\documentclass[prl,twocolumn]{revtex4}  
\usepackage{graphicx}
\usepackage{amssymb}

\def\d#1{#1^\dagger}
\def\bra#1{\langle #1|} 
\def\ket#1{|#1\rangle}

\begin{document}

\title{Sub-shot-noise heterodyne polarimetry}
\author{Sheng Feng}
\author{Olivier Pfister}
\email{opfister@virginia.edu}
\affiliation{Department of Physics, University of Virginia, 382 McCormick Road, Charlottesville, VA 22904-4714, USA}
\begin{abstract}
We report the experimental demonstration of a heterodyne polarization rotation measurement with a noise floor 4.8 dB below the optical shot noise, by use of the classically phase-locked quantum twin beams emitted above threshold by an ultrastable type-II Na:KTP CW optical parametric oscillator. We believe this is the largest noise reduction achieved to date on optical phase-difference measurements.
\end{abstract}
\maketitle

Heisenberg-limited measurements are of considerable interest for ultrasensitive interferometry.\cite{hli} Based on bosonic or fermionic quantum interferometry, such measurements possess an ultimate sensitivity limit of $1/n$ (the Heisenberg limit, HL), for a detected photon number $n$, as opposed to the beam-splitter's shot noise limit (SNL) $1/\sqrt n$. In quantum optics, the SNL applies whenever the light impinging on the two sides of the input beam splitter of an interferometer comprises one vacuum (``unused") mode.\cite{caves80} However, if this vacuum mode is replaced with nonclassical light, such as squeezed vacuum\cite{caves81} or a photon-number correlated (twin) beam,\cite{holland} the SNL can be exceeded and, for ideal (infinite) squeezing, the measurement becomes Heisenberg-limited. Sub-SNL measurements have been demonstrated using vacuum squeezing,\cite{xiao,grangier} twin-photon pairs,\cite{kuz} and twin pulsed beams,\cite{silb} all of which involving {\em frequency degenerate} two-beam inputs into an interferometer. 

\begin{figure}[htb]
\begin{center}
\begin{tabular}{c}
\includegraphics[width=3in]{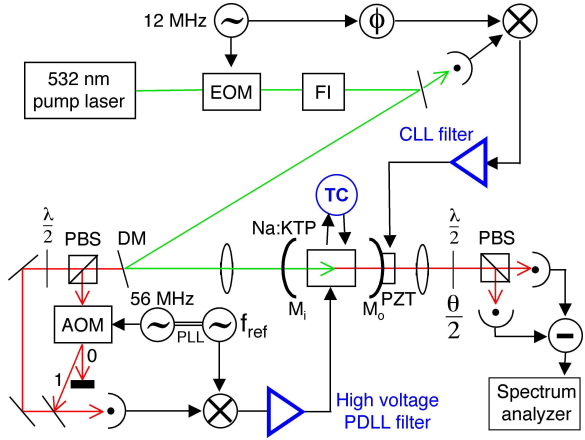}
\end{tabular}
\end{center}
\caption{Experimental setup. $\rm M_i$: input mirror (reflectivity: 0\% @ 532 nm; 99.99\% @ 1064nm). $\rm M_o$: output mirror (reflectivity: 99.995\% @ 532 nm; 98\% @ 1064nm). FI: Faraday isolator. EOM: electro-optic modulator. DM: dichroic mirror. PBS: polarizing beam splitter. AOM: acousto-optic modulator. PLL: electronic phase-lock loop. PZT: piezoelectric transducer.}
\label{setup}
\end{figure}

Another interesting Heisenberg-limited measurement, proposed by Snyder {\em et al.}, is to use frequency {\em nondegenerate} twin beams\cite{tb} for ultrasensitive polarization rotation measurements.\cite{snyder} The principle of the method is the following: a type-II optical parametric oscillator (OPO) above threshold emits intense, orthogonally polarized, photon-number correlated (twin) beams at respective frequencies in narrow bands around $\nu_{1,2}$. A photodetector of bandwidth greater than (or centered at) $\nu_-=|\nu_1-\nu_2|$ is placed at the OPO ouput. At first, no AC signal is observed as the two orthogonally polarized OPO waves cannot interfere. However, if a polarization-rotating medium is inserted, followed by a polarizer aligned with the OPO crystal axes (Fig.\ \ref{setup}), then the photodetection signal displays a heterodyne oscillation at $\nu_-$ (or beat note) due to the interference of the OPO waves. This signal increases with the polarization rotation angle. Such a heterodyne measurement can be made at high frequencies and can be as narrowband as the beat note linewidth. Its signal-to-noise ratio is high since classical (e.g.\ $1/f$) noise rolls off at the detection frequency and narrow detection bandwidth yields a reduced noise floor. It is well known that spectacular improvements can be obtained in ultra-precise measurements by implementing classical heterodyne techniques.\cite{fm} Thus, heterodyne measurements are often shot-noise limited and quantum optical techniques, such as squeezing, can then be applied to further reduce the noise floor. Going back to our case, Snyder {\em et al.}\ predicted that, because of the photon-number correlation of the twin beams within the OPO cold-cavity linewidth $\gamma$, the polarization rotation beat note should rest on a sub-shot noise floor if $\nu_-<\gamma$. For this to occur, the rotation angle $\theta$ must be small enough that the resulting beam mixing not destroy the quantum intensity correlation.\cite{fengPRL} Such a method is therefore well suited to the ultrasensitive detection of minute birefringence. 
This Letter presents the experimental realization of sub-SNL heterodyne polarimetry. 

We first show that this measurement is, ideally, Heisenberg-limited. One could indeed wonder whether the HL itself would not be breached by Snyder's scheme since the ideal twin-beam noise floor is zero, not $1/n$. (The photon number difference is a constant of motion whose initial state is a twin vacuum state.) However, a polarization rotation $\theta\neq 0$ yields, in fact, no heterodyne signal at all if the twin beam correlation is ideal. To see this, consider the two outputs of the aforementioned polarizer $c= a_1\cos\theta+ b_2\sin\theta$ and $d= a_1\sin\theta- b_2\cos\theta$, where $a_1$, $b_2$ are the twin input modes at frequencies $\nu_{1,2}$. The number difference is $\d cc-\d dd=\cos 2\theta(\d a_1a_1-\d b_2b_2)+\sin 2\theta(\d a_1b_2+a_1\d b_2)$, the second term being the beat note. If one considers a general ideal twin-beam state $\ket{\psi}=\sum_n c_n\ket n_a\ket n_b$ (i.e.\ a zero noise floor) then the average signal $\bra{\psi}\d cc-\d dd\ket{\psi} = 0$, $\forall \theta$, since $\bra{n' n'}\d a_1b_2\ket{n n} \propto\delta_{n'\,n+1}\delta_{n'\,n-1}$. It is therefore necessary, for a given photon number $n$,  to have deviations from perfect photon-number correlation by at least $\pm 1$ photon in order to observe a nonzero beat note signal $\sim n\sin 2 \theta$. Hence, the smallest detectable angle is $\theta_{\mathit min}\sim\pm 1/(2n)$, of the order of the HL.

In practice, two technical problems must be solved. The first one is the residual classical frequency jitter of the beat note, which can mask the noise reduction if the bandwidth of the latter is comparable to the jitter excursion range. Recently, our group has achieved above-threshold nondegenerate OPO operation at ultrastable frequency differences (and orthogonal polarizations) in a routinely repeatable way, by electronic servo stabilization of the phase difference of the twin beams.\cite{fengSPIE} This lead to the observation of generalized Hong-Ou-Mandel interference between interfering states with photon numbers much larger than one,\cite{fengPRL} and paves the way towards a realization of continuous-variable entanglement of ultrastable bright beams.\cite{reid} When the beat note is provided by our phase-locked OPO (whose frequency difference is ultrastable yet continuously tunable), its jitter is essentially zero.\cite{fengPRL} The second problem is the existence of a large residual beat note, even when $\theta=0$, because of polarization crosstalk in the OPO cavity. If not carefully canceled, this residual beat note can reach a few percent of the maximum amplitude and its pedestal can then easily overwhelm the noise floor. It is thus crucial to suppress this residual beat note. It is known that polarization crosstalk in a type-II OPO is linked to critical phasematching and walkoff in the OPO crystal.\cite{ou} In order to suppress polarization crosstalk, we  noncritically phasematch frequency degenerate parametric oscillation at 1064 nm, pumped at 532 nm, by using a Na:$\rm KTiOPO_4$ (Na:KTP) crystal.\cite{fengJOB} By carefully aligning the OPO so that the cavity axis coincide with the $X$ principal axis of the crystal,we can obtain a reduction of the residual beat note by more than 65 dB. However, other experimental factors such as wave plate and polarizer imperfections will also yield a beat note which is just the illustration of the {\em classical} sensitivity of heterodyne polarimetry using the perfectly mode-matched OPO beams. We will therefore not worry about the residual beat note, now but a technical issue as it is small enough for the quantum noise floor to be well resolved. 

The experimental setup is depicted in Fig.~\ref{setup}.
The OPO consists of a Na:KTP nonlinear crystal, stabilized by a temperature controller (TC) at the 0.1 mK level, in which pump photons at 532 nm are downconverted to cross-polarized pairs at 1064 nm. The OPO cavity is formed by mirrors $\rm M_i$ and $\rm M_o$. The 1064 nm twin beams exit through $\rm M_o$ to the right of the figure. The reflected depleted pump beam provides the error signal for the cavity lock loop (CLL), and a weak twin-beam leak through $\rm M_i$ is used for the phase-difference lock loop (PDLL). With only the TC and CLL on, the frequency difference error is $\pm$150 kHz, due to large Na dopant inhomogeneities in the crystal (the Na proportion is 23\%).\cite{fengJOB} Adding the PDLL reduces this error by more than 5 orders of magnitude to less than 1 Hz,\cite{fengPRL,fengSPIE} while keeping the frequency difference continuously temperature-tunable over tens of MHz. Finally, since the PDLL error signal is obtained from beams leaking through a mirror with very low transmission, it is entirely classical and the PDLL cannot modify the quantum phase fluctuations. 

\begin{figure}[htb]
\begin{center}
\begin{tabular}{c}
\includegraphics[width=3.5in]{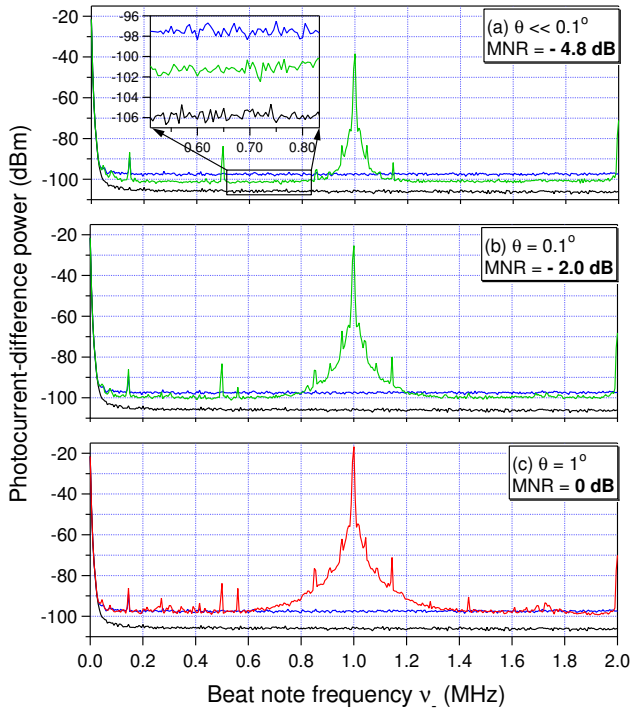}
\end{tabular}
\end{center}
\caption{Sub-SNL heterodyne polarimetry signals. In all three figures, the two flat traces are the shot noise level (upper) and the detection electronics noise (lower). The peaked trace is the twin-beam beat note signal, whose subhertz beat note linewidth\cite{fengPRL,fengSPIE} is not directly visible on these records since the spectrum analyzer's resolution and video bandwidths are 3 kHz (100 averages). The maximum beat note amplitude ($\theta=\rm 22.5^o$) is 25 dBm. MNR: Measurement-noise reduction.}
\label{hp}
\end{figure}

The OPO is operated at pump beam powers a few percent above the 50 mW threshold. The twin beams, each of power 1.4 mW, exit the cavity through $\rm M_o$ and traverse a half-wave plate that can be rotated by $\theta/2$, followed by a polarizing beam splitter aligned with the OPO polarizations. The polarizer is initially aligned without the wave plate so that the residual beat note at $\nu_-$ is minimized. Then the wave plate is inserted and an increase in the beat note is observed, even if $\theta=p\frac{\pi}2$, due to  imperfections. The beat note is temperature-tuned to 1 MHz, well within the twin-beam squeezing bandwidth of 5 MHz, and phase-locked. The rotation angle is then increased. Figure \ref{hp} displays the photocurrent difference power spectrum for $\theta\rm \ll 0.1^o$, $\theta\rm \simeq 0.1^o$, and $\theta\rm \simeq 1^o$. (The beat note contrast is almost unity due to the near ideal mode-matching of the twin beams coming out of the high-finesse OPO resonator.)
The sub-shot-noise measurement background is clearly visible for small polarization rotations in Figs.\ \ref{hp} (a,b), illustrating the polarimetry sensitivity increase. The maximum noise-floor reduction is $-4.8$ dB in Fig.~\ref{hp} (a) and inset (taking the electronic noise floor into account but not the 94\% quantum efficiency of the detectors). Note that the squeezing occurs at frequencies as low as 100 kHz. The beat note has a noise pedestal which, we surmise, is related to the Schawlow-Townes phase-difference noise of the OPO,\cite{jdp} such as previously observed by us for frequency-degenerate twin beams (0 Hz beat note).\cite{fengPRL} As $\theta$ increases from $\rm 0^o$ up to $\rm 1^o$, this noise pedestal grows with the beat note and eventually overwhelms the squeezed background. This noise increase could be caused by contamination by the OPO antisqueezed phase-difference noise as well as vacuum fluctuations, in particular from the ``image bands".\cite{ys} The detailed theoretical analysis of the noise spectrum of this heterodyne measurement is fairly involved and is in progress. Note that additional classical noise is also visible on the beat note envelope in the form of oscillation sidebands: these are the 22 kHz servo oscillations of the CLL, which are driven by the PDLL gain (the PDLL itself has much faster oscillations). Because of the various tuning coefficients,\cite{fengJOB} the coupling between the CLL and PDLL only occurs one-way and the PDLL is not perturbed back by the CLL.\cite{fengSPIE} 

Several factors ought to be pointed out that will significantly improve the performance. First, we are probably limited in the polarization crosstalk cancellation by the strong inhomogeneity of the Na:KTP crystal, which makes it impossible to have a clean separation of the signal and idler propagating in the crystal. Second, residual absorption in the crystal is not as low as what can be achieved in hydrothermally grown or grey-track resistant KTP. Our maximum twin beam correlation level (-7 dB) is very position-dependent in the crystal and does not match recent records in excess of -9 dB.\cite{rec} One ought to bear in mind, however, that noncritical phasematching in the OPO is indispensable here to suppress the residual polarization crosstalk and, therefore, regular KTP cannot do. A possible course of action would be to use periodically poled low-loss ferroelectrics such as hydrothermally grown or grey-track resistant PPKTP.

Finally, we compare this measurement with previous realizations of sub-SNL interferometry, using squeezed vacuum,\cite{xiao,grangier} twin-photon,\cite{kuz} and twin-pulse\cite{silb} inputs. Our measurement has the largest noise reduction to date. Moreover, in all other measurements, the two input fields are indistinguishable, in particular frequency-degenerate, whereas here the fields are frequency nondegenerate and clearly distinguishable. The heterodyne nature of the signal thus makes it easy to escape classical and $1/f$ noise. Because of the exquisite degree of control that our ultrastable CW OPO displays, we believe further improvement of our measurements beyond -4.8 dB is well within reach.

We thank Ken Nelson for precious electronics design advice, Harvey Sugerman for electronics realization, Rodger Ashley and Roger Morris for machining and helping design the OPO structure, and Dan Perlov, from Coherent Crystal Associates Inc., for his help with understanding Na:KTP. This work was supported by ARO grants DAAD19-01-1-0721 and DAAD19-02-1-0104.


\begin{thebibliography}{99}
\bibitem{hli} A. Luis and L.L. S\'anchez-Soto, Progress in Optics {\bf 41}, Ed.\ E. Wolf, Elsevier (2000) p.\ 421.
\bibitem{caves80} C.M. Caves, \prl {\bf 45}, 75 (1980).
\bibitem{caves81} C.M. Caves, \prd {\bf 23}, 1693 (1981).
\bibitem{holland} M.J. Holland and K. Burnett, \prl {\bf 71}, 1355 (1993).
\bibitem{xiao} M. Xiao, L.-A. Wu, and H.J. Kimble, \prl {\bf 59}, 278 (1987).
\bibitem{grangier} P. Grangier, R.E. Slusher, B. Yurke, and A. LaPorta, \prl {\bf 59}, 2153 (1987).
\bibitem{kuz} A. Kuzmich and  L. Mandel, Quantum Semiclass.\ Opt.\ {\bf 10}, 493 (1998). 
\bibitem{silb} Ch.\ Silberhorn, P. K. Lam, O. Wei\ss, F. K\"onig, N. Korolkova, G. Leuchs, \prl {\bf 86}, 4267 (2001).
\bibitem{tb} A. Heidmann, R.J. Horowicz, S. Reynaud, E. Giacobino, and C. Fabre, \prl {\bf 59}, 2555 (1987).
\bibitem{snyder} J.J. Snyder, E. Giacobino, C. Fabre, A. Heidmann, and M. Ducloy, \josab {\bf 7}, 2132 (1990).
\bibitem{fm} G.C.~Bjorklund, \ol {\bf 5}, 15 (1980); J.L.~Hall, L. Hollberg, T. Baer, and H.G. Robinson, \apl {\bf 39}, 680 (1981).
\bibitem{fengPRL} S. Feng and O. Pfister,  \prl {\bf 92}, 203601 (2004).
\bibitem{fengSPIE} S. Feng and O. Pfister, Proc. SPIE {\bf 5161}, 109 (2004).
\bibitem{reid} M.D. Reid and P.D. Drummond, \prl {\bf 60}, 2731 (1988).
\bibitem{fengJOB} S.~Feng and O.~Pfister, J. Opt.\ B: Quantum Semiclass.\ Opt.\ {\bf 5}, 262  (2003). 
\bibitem{ou} Z.Y. Ou, S.F. Pereira, H.J. Kimble, and K. Peng, \prl {\bf 68}, 3663 (1992).
\bibitem{jdp} C. Fabre, E. Giacobino, A. Heidmann, S. Reynaud, J. Phys.\ {\bf 50}, 1209 (1989). 
\bibitem{ys} H.P. Yuen and J.H. Shapiro, IEEE Trans.\ Inform.\ Theory {\bf IT-26}, 78 (1980).
\bibitem{rec} J. Gao, F. Cui, C. Xue, C. Xie, and K. Peng, \ol {\bf 23}, 870 (1998); J. Laurat, T. Coudreau, N. Treps, A. Ma\^itre, and C. Fabre, \prl {\bf 91} 213601, (2003).
\end{thebibliography}
\end{document}